# Coherent spin valve phenomena and electrical spin injection in ferromagnetic/semiconductor/ferromagnetic junctions

Francisco Mireles[(1)] and George Kirczenow[(2)]
[1]*Centro de Ciencias de la Materia Condensada UNAM, 22800 Ensenada BC, México.*
[2]*Department of Physics, Simon Fraser University, Burnaby, BC, Canada, V5A 1S6.*

Coherent quantum transport in ferromagnetic/ semiconductor/ ferromagnetic junctions is studied theoretically within the Landauer framework of ballistic transport. We show that quantum coherence can have unexpected implications for spin injection and that some intuitive spintronic concepts which are founded in semi-classical physics no longer apply: A *quantum spin-valve* (QSV) effect occurs even in the *absence* of a net spin polarized current flowing through the device, unlike in the classical regime. The converse effect also arises, *i.e.* a *zero* spin-valve signal for a non-vanishing spin-current. We introduce new criteria useful for analyzing quantum and classical spin transport phenomena and the relationships between them. The effects on QSV behavior of spin-dependent electron transmission at the interfaces, interface Schottky barriers, Rashba spin-orbit coupling and temperature, are systematically investigated. While the signature of the QSV is found to be sensitive to temperature, interestingly, that of its converse is not. We argue that the QSV phenomenon can have important implications for the interpretation of spin-injection in quantum spintronic experiments with spin-valve geometries.

PACS numbers: 72.25.Hg, 73.40.Sx, 72.25.Mk

## I. INTRODUCTION

Electrical spin-injection of coherent polarized carriers from ferromagnetic metals into semiconductors is currently an issue of fundamental relevance in spintronics.[1,2]It was suggested in the seminal work of Datta and Das[3] that the use of ferromagnetic metals as source and drain contacts (spin-injector and -detector) connected to a semiconductor would make feasible a unique transistor that relies on the manipulation of the electron's spin instead of its charge. Since then considerable effort has been directed towards practically demonstrating efficient injection of spin-polarized electrons through ferromagnetic/semiconductor (F/S) solid state interfaces.[4–9] This issue has been one of the central challenges in the field, as its demonstration have encountered crucial obstacles, such as the large resistivity mismatch of typical metals and semiconductors, a condition that severely inhibits spin-injection.[10,11]

Work is in progress to determine whether these obstacles may be overcome through the use of suitable potential barriers[11–13] or through appropriate epitaxial interfaces that obey certain selection rules and band structure symmetry properties,[14] as recently corroborated by *ab initio* spin-transport calculations.[15–17] For instance, recently, Hammar and Johnson[18] have performed successful spin-dependent transport measurements across ferromagnetic-metal/insulating barrier/two-dimensional electron gas (2DEG) junctions, validating the theoretical predictions.[11–13] Recent experiments at room temperature on spin injection from ferromagnetic metal contacts into a semiconductor (Fe/GaAs) via STM through Schottky tunnel barriers[19,20] have yielded encouraging results of about 2% injection efficiencies, and very recently, Hanbicki *at al.*[21] have achieved relatively high spin-injection efficiencies of 30% in Fe/GaAs-based light emitting diode structures, showing the effectiveness of the Fe Schottky tunnel contacts in enhancing the spin efficiency rates.[21,22]

These experiments suggest that combinations of ferromagnetic metals and semiconductor materials may be indeed promising for applications in hybrid semiconductor spintronic devices. It should also be noted that very high electron (or hole) spin-injection efficiencies have been achieved from magnetic to non-magnetic semiconductors. However this has required low temperatures and/or strong magnetic fields.[23,24]

Most of the theoretical modeling of spin dependent transport in two-terminal F/S/F systems reported to date has been in the semi-classical diffusive regime of transport (within the Boltzmann framework),[10–13,25–27] with just a few studies in ballistic regime.[28–30] However, it has been pointed out that quantum interference phenomena may be exploited in novel spintronic devices.[31–33] Quantum effects on the spin injection such as quantum coherence and interference have been typically neglected until recently. The interference effects in a F/S/F transistor were studied by Schäpters *et al.*[34] showing that an enhanced spin signal can be attained when quantum interference is considered. Also recently, Matsuyama *et al.*[35] have studied ballistic spin transport in ferromagnet/2DEG/ferromagnet double junctions taking into account the spin-orbit interaction in the quantum ballistic regime. In a more recent work we have explored the interplay between spin injection and quantum coherence in ballistic F/S/F heterojunctions theoretically within the Landauer formalism of transport.[36]

We showed that quantum coherence give rise to a *Quantum Spin Valve* (QSV) effect that, unlike its familiar semi-classical analog, occurs even in the absence of a net spin current flowing through the heterostructure.[37]

The purpose of this work is to provide a comprehensive and detailed study of the QSV effect in F/S/F junctions, and investigate theoretically to what extent the effects of quantum interference,[28] spin-dependent electron transmission at the interfaces,[29] interface Schottky barriers,[34,35] Rashba spin-orbit coupling,[38–40] and temperature effects are of relevance for the existence on the predicted *quantum* spin-valve behavior. We establish new criteria that are helpful in interpreting and analyzing quantum and classical spin transport phenomena. It is then verified that the QSV effect is an inherently quantum interference process and we find that its distinctive signature is extremely sensitive to temperature. However the converse of the QSV, although it is also due to quantum interference, is found to be remarkably temperature-insensitive. We find that the QSV effect persists even in the presence of Schottky barriers at the boundaries and that these enhance spin-injection in the quantum coherent regime, although rather weakly. Our results show that, in the *ballistic quantum regime* of transport care has to be exercised in order to appropriately interpret the physics of spin injection experiments with spin-valve geometries.

The paper is organized as follows: In Sec. II we establish and discuss the semiclassical picture of electron spin-injection at F/S interfaces. In Sec. III the classical spin-valve concepts are outlined, their connection with spin-injection in F/S/F junctions is discussed, and new criteria are established for interpreting both classical and quantum spin transport behavior. Sec. IV is devoted to the description of the ballistic quantum approach that we use to model the spin-transport mechanism. The results and our discussion of them are given in Sec.V. Finally in Sec. VI a summary and conclusions are presented.

## II. SPIN INJECTION RATE

It is instructive to first review the standard (classical) definition of the electron spin-injection rate at a single F/S interface. This will help us to establish the fundamental physical concepts that we will use later to examine the case of spin-injection in a F/S/F heterojunction. Afterwards we will describe how the spin-valve phenomenon (a change in the resistance when the magnetization of a ferromagnetic electrode is reversed) is related to spin-injection in two terminal devices in the semi-classical picture.

Following Johnson and Silsbee,[26] consider an ideal Stoner-Wohlfarth model for a ferromagnetic metal with just one parabolic spin-subband at the Fermi surface, and in equilibrium with a semiconductor material, whose (two-fold degenerate) spin subband structure is assumed to be free-electron like. By applying a potential bias $V$ across the interface it is expected that electrons of that spin-subband will be driven into the semiconductor. Neglecting spin-relaxation at the interface, the magnetization current transfered through the interface into the semiconductor would be proportional to the total electronic current. Since each carrier transports a spin magnetic moment with magnitude $|\vec{\mu}_B| = \mu_B$, $\mu_B = e\hbar/2mc$ being the Bohr magneton. The net injected magnetization current $j_M$ related to the driven electric current $j_e$ will be $j_M = \mu_B j_e/e$. In practice the Fermi surfaces of most ferromagnetic metals have both spin-subbands occupied, although with a significant imbalance in the density of states at the Fermi energy. Therefore the spin-injection (spin magnetization) is reduced correspondingly. Assuming weak coupling between spin-subbands, this reduced magnetization injection efficiency can be described by the dimensionless (phenomenological) parameter $\eta_M$, such that $j_M = \eta_M \mu_B j_e/e$. Explicitly, the spin-injection rate through a single F/S heterojunction is related to the electric current through the ratio[11,26]

$$\frac{j_M}{j_e} = \frac{j_\uparrow - j_\downarrow}{j_\uparrow + j_\downarrow}\frac{\mu_B}{e} \equiv \eta_M \frac{\mu_B}{e} \qquad (1)$$

The interfacial transport parameter $\eta_M$ in (1) thus describes the degree of spin-polarization of the net electron flux through the interface. Note that $|\eta_M| \leq 1$. We notice also that generally speaking, as was pointed out by Johnson and Silsbee[11], the current of non equilibrium magnetization may be written as $\overleftrightarrow{j}_M = \eta(\vec{\mu}_M/e)\vec{j_e}$. Here $\overleftrightarrow{j}_M$ is a second-rank tensor which specifies both the direction of flow and the orientation of the magnetization, such that the component $(\overleftrightarrow{j}_M)_{\alpha\beta}$ describes the transport along the $\beta$-axis of the projection of the magnetization on the $\alpha$-axis.[11,26] Assuming that the charge current is along the $x$-axis, normal to the interface, that is $\vec{j_e} = j_e\hat{x}$, and the magnetization of the ferromagnet metal is such that the spin-polarization is aligned along the $z$-axis, for instance, then $\overleftrightarrow{j}_M = \eta(\mu_M/e)j_e\hat{z}\hat{x} = j_M\hat{z}\hat{x}$, therefore $\overleftrightarrow{j}_M$ and $\vec{j_e}$ can be treated simply as scalars.[26]

The ratio (1), was originally introduced for ferromagnetic metal/paramagnetic metal interfaces, but applies equally to F/S interfaces. Within the linear response regime assuming that no spin-flip scattering at the interface[26] or spin-precession[39] is present, and in terms of the spin-conductances $G_\sigma$, we have $j_e = j_\uparrow + j_\downarrow = (G_\uparrow + G_\downarrow)V$. Therefore

$$\eta_M = \frac{G_\uparrow - G_\downarrow}{G_\uparrow + G_\downarrow}. \qquad (2)$$

This relationship clearly shows that there is a net flux of spin-polarized electrons through the F/S interface for all $G_\uparrow \neq G_\downarrow$.

Equation (2) can be extended to F/S/F double interface heterojunctions. Then, in terms of the total spin-conductances for the entire device (assuming that the ferromagnetic contacts have parallel magnetization) the spin injection efficiency is given by

$$\eta'_M = \frac{G^{tot}_\uparrow - G^{tot}_\downarrow}{G^{tot}_\uparrow + G^{tot}_\downarrow}. \tag{3}$$

In the semi-classical regime of transport where all quantum phase information is assumed to be lost during (ballistic) electron transit between interfaces (that is, by neglecting all phase information in the calculation of net transmission through two scatterers in series) the elastic multiple scattering at the interfaces results in the following spin-transmission probabilities[29]

$$T^{tot}_\sigma = T_{P\sigma} = \frac{T_\sigma}{2 - T_\sigma}, \tag{4}$$

where $T_\sigma$ are the independent single-interface transmission probabilities, with spin $\sigma = (\uparrow, \downarrow)$, and $P$ denotes the parallel configuration of the ferromagnets. Hence (semi-classically), from (4) with $G^{tot}_\sigma = (e^2/h)T^{tot}_\sigma$ within the ballistic Landauer picture,[36] the spin-injection efficiency in Eq.(3) for the F/S/F structure can be written as

$$\eta'_M = \frac{T_\uparrow - T_\downarrow}{T_\uparrow + T_\downarrow - T_\uparrow T_\downarrow}, \tag{5}$$

which in turns suggests that a net spin current flows across the F/S/F heterojunction whenever $T_\uparrow \neq T_\downarrow$.

A particularly important spintronic phenomenon somewhat related to the injection of spin-polarized currents is the magnetoconductance (magnetoresistance), also dubbed spin-valve behavior. In the following section we shall discuss the relationship between the widely accepted definition of magnetoconductance and spin-injection of polarized electrons in two terminal structures in the semi-classical picture. We will then examine to what extent this relationship may be extended to the case of a coherent quantum regime of transport.

## III. SPIN VALVE PHENOMENA

The absence of a complete theory of spin-injection in the ballistic quantum regime of transport, has led in part, to borrowing criteria for the efficency of spin-injection from the semi-classical ballistic and diffusive approaches of spin-transport. Thus a well known, although *indirect* way to electrically detect spin-injection experimentally is based on the *spin-valve* effect, a phenomenon known to be yielded by multiple spin-dependent electron scattering events at the interfaces of ferromagnetic/non-magnetic junctions. It is a measure of the change in conductance (or resistance) when the magnetizations of the ferromagnetic contacts in a F/SP/F switch between the parallel (P) and anti-parallel (AP) configurations. Here SP stands for a spacer which can be a normal metal, a semi- or super- conductor. This change is normally represented by the ratio[9,34,35,41]

$$\frac{\Delta G}{2G_{av}} = \frac{\Delta R}{2R_{av}} = \frac{G^P - G^{AP}}{G^P + G^{AP}} \equiv \eta. \tag{6}$$

Hence $\eta$ can be seen as the normalized change in conductance between the parallel and anti-parallel configurations of the magnetic moments of the two ferromagnetic electrodes. Now, in the semi-classical picture, where the interfaces are simply regarded as elastic phase-incoherent scatterers (resistors) in series, the transmission probabilities per spin orientation for the anti-parallel configuration are given by

$$T_{AP\uparrow} = T_{AP\downarrow} = \frac{T_\uparrow T_\downarrow}{T_\uparrow + T_\downarrow - T_\uparrow T_\downarrow}, \tag{7}$$

with the total transmission given by $T_{AP} = T_{AP\uparrow} + T_{AP\downarrow}$. Hence it follows that the relative conductance or magnetoconductance ratio can be written also in terms of the single interface spin-probabilities,

$$\eta = \frac{(T_\uparrow - T_\downarrow)^2}{T^2_\uparrow + 6T_\uparrow T_\downarrow - 4T^2_\uparrow T_\downarrow - 4T_\uparrow T^2_\downarrow + 2T^2_\uparrow T^2_\downarrow + T^2_\downarrow} \tag{8}$$

Thus semiclassically, as for $\eta'_M$, $\eta$ is not zero when $T_\uparrow \neq T_\downarrow$. Thus (in geometries that exclude extrinsic signals due to local Hall fields and the like[9]) the observation of a spin-valve effect ($\eta \neq 0$) in the semi-classical ballistic regime implies that spin-injection is taking place and vice versa. Similarly, it is generally believed (with the same caveat[9]) that observation of a spin valve effect in the semi-classical diffusive regime indicates that spin injection is taking place and that the same is true for all-metal systems.

Interestingly enough, we shall see below that the above semi-classical, and somewhat intuitive arguments for the interpretation of the magnetoconductance ratio $\eta$, as a measure of a spin-injection rate, does not necessarily hold in the ballistic quantum coherent regime of transport.

In order to better understand the condition(s) for the occurrence (or absence) of an overall spin-injection in a F/SP/F heterojunction, we will rewrite the expression (6) for the relative magneto conductance in a slightly different way and in terms of the net spin-currents flowing through the heterojunction. This will allow us to have a clearer physical insight to the issue of the correctness of the interpretation for $\eta$, when trying to elucidate its physical significance in the ballistic quantum regime of transport, where quantum coherence and interference can play a fundamental role.

We start by noticing that, since $\Delta j^{AP} = j^{AP}_\uparrow - j^{AP}_\downarrow = 0$ as long the right and left ferromagnets are of the same material, and no external magnetic fields are present, then the (spin-valve) magnetoconductance coefficient

$$\eta = \frac{G^P - G^{AP}}{2G_{av}} = \frac{j^P - j^{AP}}{2j_{av}}, \tag{9}$$

where the subscript $(av)$ denotes the average value between the parallel and antiparallel conductance (or current), can without loss of generality, conveniently be rewritten as

$$\eta = \frac{\Delta j^P - 2(j_\uparrow^{AP} - j_\downarrow^P)}{j^P + j^{AP}}, \tag{10}$$

where $\Delta j^P = j_\uparrow^P - j_\downarrow^P$ represents the net electron spin-current (magnetization) flowing through the device in the parallel (P) configuration. The term $2(j_\uparrow^{AP} - j_\downarrow^P)$ does not have, an apparent physical meaning, since it depends on independent conductances (current) measurements, at least in the semi-classical picture. However we will show below that it can play an important role in the quantum regime.

Clearly from (10) a null result in the magnetoconductance ($\eta = 0$) will imply either of the following conditions (omitting the obvious case $j_\uparrow^{P/AP} = j_\downarrow^{P/AP} = 0$):

($a.1$) $\Delta j^P = 0$ and $2(j_\uparrow^{AP} - j_\downarrow^P) = 0$ , or

($a.2$) $\Delta j^P = 2(j_\uparrow^{AP} - j_\downarrow^P) \neq 0$.

On the other hand, the situation with finite magnetoconductance, $\eta \neq 0$, should always occur whenever:

($b.1$) $\Delta j^P \neq 2(j_\uparrow^{AP} - j_\downarrow^P) \neq 0$, or
($b.2$) $\Delta j^P \neq 0$ and $2(j_\uparrow^{AP} - j_\downarrow^P) = 0$ , or
($b.3$) $\Delta j^P = 0$ and $2(j_\uparrow^{AP} - j_\downarrow^P) \neq 0$.

We emphasize that Eq. (10) and the criteria ($a.1$), ($a.2$), ($b.1$), ($b.2$), and ($b.3$) that follow from it are quite general and apply to both classical and quantum systems.

Let us now examine in detail the consequences of these conditions. In the ballistic *semi-classical* regime, if $\Delta j^P = 0$ then from (4) $T_\uparrow = T_\downarrow$ since $T_{P\uparrow} = T_{P\downarrow}$, and therefore it follows from (4) and (7) that $T_{AP\uparrow} = T_{P\downarrow} = T_{AP\downarrow} = T_{P\uparrow}$, and hence $2(j_\uparrow^{AP} - j_\downarrow^P) = 0$. Thus, semi-classically the condition ($a.1$) is clearly fulfilled whenever $\Delta j^P = 0$. Therefore, semi-classically the *absence* of a net spin-injection ($\Delta j^P = 0$) implies that the magnetoconductance $\eta = 0$. The condition ($a.2$) can give rise to a rather misleading interpretation if it is not analyzed appropriately. It implies that the magnetoconductance $\eta$ can in fact be *zero*, but with a *nonzero* spin current ($\Delta j^P \neq 0$) flowing through the device. There is an apparent inconsistency here, since it would appear to contradict the intuitive standard criteria for the existence of spin-injection, *i.e.* $\eta = 0 \rightarrow \Delta j^P = 0$ (or $T_{P\uparrow} = T_{P\downarrow}$). However, this condition ($a.2$) never occurs in the *semi-classical* regime as the only way to satisfy $\Delta j^P = 2(j_\uparrow^{AP} - j_\downarrow^P)$ in that regime is to have $\Delta j^P = 0$, ($T_\uparrow = T_\downarrow$) which brings us back to condition ($a.1$). Hence the measurement of a *zero* magnetoconductance will ensure that spin-injection is not taking place, at least in the ballistic semi-classical regime of transport. Nevertheless, it is clear that the occurrence of $\eta = 0$ may *in principle* be allowed for a non vanishing $\Delta j^P$. We will see below that this can in fact occur in the *coherent quantum* regime of transport, leading thus to counterintuitive results if one tries to interpret them within the framework of the semi-classical criteria of spin-injection.

On the other hand, a measurement of $\eta \neq 0$ likewise has interesting consequences as we seek again to interpret its significance in relation to spin-injection. Note for instance that $\Delta j^P \neq 0$ in both conditions ($b.1$) and ($b.2$), which in turn physically implies a finite spin-current. Therefore it becomes evident that the interpretation of the condition $\eta \neq 0$ as a criterion indicative of the presence of a net spin-injection is always valid, except in the case ($b.3$). In the case ($b.3$) we have *zero* spin-current, which would appear to contradict the criterion that $\eta \neq 0$ implies finite spin-injection. However, in the semi-classical regime the situations $\Delta j^P = 0$ and $2(j_\uparrow^{AP} - j_\downarrow^P) \neq 0$ are never both satisfied at the same time (see Eqs. (4) and (7)), and consequently ($b.3$) never occurs in this regime. Similarly, it can be shown that condition ($b.2$) never holds in this regime, despite $\Delta j^P$ being *nonzero*. In other words, in the semi-classical picture, for all $\eta \neq 0$, only the condition ($b.1$) is fulfilled, and there is no possibility that a physically counter-intuitive situation will occur. Therefore, the observation of a *nonzero* magnetoconductance ($\eta \neq 0$ ) by itself consititutes unequivocal evidence of a net spin-current injection in the semi-classical regime. However, the possibility of having a non vanishing $\eta$ without having any spin-current at all ($b.3$) is in principle conceivable and in that case the *classical* interpretation of the criteria for spin-injection break down. Indeed this situation is realized in the quantum regime giving rise to a *quantum spin-valve* effect, that we shall describe below. Therefore special care has to be exercised in the ballistic quantum regime of transport for the appropriate interpretation of spin-injection experiments.

We now turn to the description of the quantum coherent spin-transport model in a F/S/F heterostructure that we will use to make a systematic study the behavior of the spin-injection in this regime in spin-valve systems.

## IV. SPIN TRANSPORT MODEL: QUANTUM REGIME

We consider ballistic spin transport through a F/S/F hybrid heterojunction. In the (identical) ferromagnetic electrodes a Stoner-Wohlfarth like model of the magnetization is assumed such that the spin-up and spin-down band energies offset is set by an exchange splitting $\Delta$ (Fig. 1). The electrode magnetization is chosen along the $z$-direction, parallel to the interface. We assume the semiconductor region to have a quasi-one dimensional wave guide shape which laterally confines the electrons in the direction transverse to transport, which is assumed to be normal to the interface and along the $x$- axis. In the semiconductor channel a Rashba spin-orbit coupling[38] widely believed to be of importance in narrow gap semiconductors, will be also considered.[43,44] In order to incorporate the tunnel Schottky barriers usually present at F/S interfaces[34,35,41], simple delta type interface potentials are also included in our model.

The total one-electron effective mass Hamiltonian for parallel (P) magnetization of the ferromagnets is given by the sum

$$\hat{H} = \hat{H}_o + \hat{H}_{so} + \hat{H}_z + V(x), \tag{11}$$

with

$$\hat{H}_o = \frac{1}{2}\hat{p}_x \frac{1}{m^*(x)}\hat{p}_x \tag{12}$$

$$\hat{H}_{so} = \frac{1}{2\hbar}\sigma_z[\hat{p}_x \alpha_R(x) + \alpha_R(x)\hat{p}_x], \tag{13}$$

$$\hat{H}_z = \frac{1}{2}\Delta\sigma_z + (\delta E_c - \frac{1}{2}\Delta\sigma_z)\theta(x)\theta(l_s - x), \tag{14}$$

and

$$V(x) = \begin{pmatrix} V_\uparrow(x) & 0 \\ 0 & V_\downarrow(x) \end{pmatrix}. \tag{15}$$

Clearly $\hat{H}_o$ is due to the free-electron part, $\hat{H}_{so}$ introduces the Rashba spin-orbit interaction, $\alpha_R(x)$ being the position dependent spin-orbit (Rashba) parameter.[35,41,42] $\hat{H}_z$, describes the exchange interaction in the ferromagnetic metals, as well as the band-offset between the semiconductor and ferromagnet band structure at the interface, with $\delta E_c$ modeling the F/S conduction band structure mismatch. The last term, $V(x)$ defines Schottky delta barrier potentials at the interfaces which are modeled by $V(x) = U^L_\sigma \delta(x) + U^R_\sigma \delta(x - l_s)$.[34,35] Although the strength of the $\delta$-potentials $U^{L,R}_\sigma$, have been set spin-dependent for completeness, in the actual calculations concerned here they will be assumed spin-independent. Since $\theta(x)$ defines a Heaviside step function, the F/S and S/F interfaces are located at $x = 0$ and $x = l_s$, respectively. Accordingly, the position-dependent conduction effective mass is given by $m^*(x) = m^*_f + (m^*_s - m^*_f)\theta(x)\theta(l_s - x)$, with $f$ and $s$ indicating the ferromagnet and semiconductor regions, respectively. Notice that we use the one-dimensional symmetrized version of the Rashba Hamiltonian,[37,41,42] and neglect inter-subband mixing which is permissible if $W << \hbar^2/\alpha_R m^*_s$, where $W$ is the width of the transverse confining potential that defines the channel.[39,45]

### A. Spin-Transport Properties

In the ferromagnetic metal contacts the energy spectrum is given by

$$E^f_\sigma(k^f_\sigma) = \frac{\hbar^2}{2m^*_f}(k^f_\sigma)^2 + \frac{1}{2}\lambda_\sigma \Delta, \tag{16}$$

where $\sigma = \uparrow, \downarrow$ labels the spin-state of the split band structure, with $\lambda_{\uparrow,\downarrow} = \pm 1$, and direction of the spin quantization along the $z-$axis. In the semiconductor there is a Rashba splitting of the dispersion which is linear in $k$, thus

$$E^s_\sigma(k^s_\sigma) = \frac{\hbar^2}{2m^*_s}(k^s_\sigma)^2 + \lambda_\sigma \alpha_R k^s_\sigma + \delta E_c. \tag{17}$$

Now, given the spin-diagonal nature of Hamiltonian (11), we consider eigenstates of the whole F/S/F structure of the form $|\Psi_\uparrow\rangle = [\psi_\uparrow(x), 0]$, and $|\Psi_\downarrow\rangle = [0, \psi_\downarrow(x)]$. The matching boundary conditions for the wave functions at the interfaces at $x_0 = 0$ and $x_0 = l_s$ are obtained by integrating $\hat{H}|\Psi_\sigma\rangle = E|\Psi_\sigma\rangle$ from $x_0 - \epsilon$ to $x_0 + \epsilon$ in the limit $\epsilon \to 0$. This yields[37,42]

$$\left(\mu\frac{\partial}{\partial x} + \epsilon_\sigma(x)\right)\psi^f_\sigma(x)|_{x=x_o} = \left(\frac{\partial}{\partial x} + i\lambda_\sigma k_R\right)\psi^s_\sigma(x)|_{x=x_o} \tag{18}$$

$$\psi^f_\sigma(x_o) = \psi^s_\sigma(x_o) \tag{19}$$

with the definitions $\mu \equiv m^*_s/m^*_f$, $\epsilon_\sigma(x=0) = 2m^*_s U^L_\sigma$, $\epsilon_\sigma(x = l_s) = -2m^*_s U^R_\sigma$, and $k_R = m^*_s \alpha_R/\hbar^2$, the Rashba spin-orbit wave vector. The largest experimental value reported to date for $\alpha_R$ in InAs-based heterojunctions is $\alpha_R = 3 \times 10^{-12} eV\, m$, which corresponds to a Rashba wave vector of $k_R = 1.5 \times 10^5 cm^{-1}$.[43] In the ferromagnetic regions the eigenstates have the general plane-wave form

$$\psi^{f,\nu}_\sigma(x) = A^\nu_\sigma e^{ik^\nu_{F\sigma}x} + B^\nu_\sigma e^{-ik^\nu_{F\sigma}x}, \tag{20}$$

with $\nu = L, R$ denoting the left and right ferromagnet electrodes. $k^\nu_{F\sigma}$ is the Fermi wave vector for the band with spin state $\sigma$ in the ferromagnet $\nu$. In the semiconductor the general solutions will be of the form

$$\psi^s_{\uparrow,\downarrow}(x) = C_{\uparrow,\downarrow} e^{ik^s_{F\uparrow,\downarrow}x} + D_{\uparrow,\downarrow} e^{-ik^s_{F\downarrow,\uparrow}x}, \tag{21}$$

where $k^s_{F\sigma}$ is the Fermi wave vector in the semiconductor for the spin-orbit-split band with spin $\sigma$. For the parallel ($P$) magnetic configuration, *i.e.*, when the orientations of the magnetic moments of the left (L) and right (R) ferromagnets are parallel [$\vec{m}_L = \vec{m}_R = (0, 0, 1)$], the spin transmission coefficients $t^P_\sigma$ are determined by using the boundary conditions (18) and (19) and applying the transfer matrix technique. The probability of an incoming electron from the left ferromagnet at the Fermi energy $E_F$ in spin state $\sigma$, and being transmitted to the right ferromagnet with parallel (P) magnetization is thus determined by

$$T^P_\sigma = \frac{v^R_{F\sigma}}{v^L_{F\sigma}} \frac{1}{|M^\sigma_{11}|^2}, \tag{22}$$

where $v^L_{F\sigma} = \hbar k^L_{F\sigma}$ and $v^R_{F\sigma} = \hbar k^R_{F\sigma}$, are the Fermi velocities of an incoming/outgoing electron with spin $\sigma$, respectively. Explicitly the transfer matrix element $M^\sigma_{11}$ reads

$$M^\sigma_{11} = \frac{e^{ik^R_{F\sigma} l_s}}{2\mu k^L_{F\sigma}(k^s_{F\uparrow} + k^s_{F\downarrow})} m^\sigma_{11}, \tag{23}$$

where for $\sigma = \uparrow$,

$$m^\uparrow_{11} = (K_s + \mu k^L_{F\uparrow} + i\xi_\uparrow(0))(K_s + \mu k^R_{F\uparrow} - i\xi_\sigma(l_s))e^{-ik^s_{F\uparrow} l_s} - (K_s - \mu k^L_{F\uparrow} - i\xi_\uparrow(0))(K_s - \mu k^R_{F\sigma} + i\xi_\sigma(l_s))e^{ik^s_{F\downarrow} l_s} \tag{24}$$

with the definition $K_s \equiv k^s_{F\sigma} + \lambda_\sigma k_R$, and with $\xi_\sigma(0) = (-2m^*_s/\hbar^2)U^L_\sigma$, and $\xi_\sigma(l_s) = (-2m^*_s/\hbar^2)U^R_\sigma$. The transmission probability for the spin state $\sigma = \downarrow$, *i.e.* $T^P_\downarrow$, is obtained from (22)-(24) through the replacement $k^{L,R}_{F\uparrow} \to k^{L,R}_{F\downarrow}$, $\xi_\uparrow(x_o) \to \xi_\downarrow(x_o)$, $k^s_{F\uparrow} \rightleftharpoons k^s_{F\downarrow}$, and $k_R \to -k_R$, respectively. Notice that energy conservation at Fermi energy requires that,

$$k^s_{F\sigma} + \lambda_\sigma k_R = \sqrt{k^2_R + \mu(k^L_{F\sigma})^2 - \frac{2m^*_s}{\hbar^2}(\delta E_c - \frac{1}{2}\lambda_\sigma \Delta)}. \tag{25}$$

For the anti-parallel magnetization ($AP$), i.e., $\vec{m}_R = -\vec{m}_L = (0, 0, -1)$, the transmission probabilities $T^{AP}_{\uparrow,\downarrow}$ are also given by Eqs. (22)-(24) with the replacement $k^R_{F\uparrow,\downarrow} \to k^L_{F\downarrow,\uparrow}$, respectively. It is clear that $T^{AP}_\uparrow = T^{AP}_\downarrow$ by symmetry as no external magnetic fields are considered. For the case of $\xi_\sigma(0) = \xi_\sigma(l_s) = 0$, that is, with no delta Schottky barriers, $U^{L,R}_\sigma = 0$, the transmission probabilities reduce to[37]

$$T^P_\sigma = \frac{4\mu^2 k^L_{F\sigma} k^R_{F\sigma}(k^s_{F\uparrow} + k^s_{F\downarrow})^2}{\kappa^2_{\sigma_+} + \kappa^2_{\sigma_-} - 2\kappa_{\sigma_+}\kappa_{\sigma_-}\cos[(k^s_{F\uparrow} + k^s_{F\downarrow})l_s]}, \tag{26}$$

with the definitions $\kappa_{\sigma_\pm} \equiv (K_s \pm \mu k^L_{F\sigma})(K_s \pm \mu k^R_{F\sigma})$, whereas $T^{AP}_\sigma$ is similarly obtained as we have argued above for the case of $U^{L,R}_\sigma \neq 0$. The spin-conductances at zero temperature are then calculated within the Landauer formalism of ballistic transport,[36] where $G^{P/AP} = (e^2/h)\sum_\sigma T^{P/AP}_\sigma$. From this, the magnetoconductance $\eta$ is then determined using Eq. (6).

We remark that because we assumed that transport is occuring in the ballistic linear response regime, calculating $\Delta j$ is exactly equivalent to evaluating $\Delta T$ for the two magnetizations (P,AP). We can thus, in a independent way determine the spin-currents from the continuity equation,[42,45,46] which leads to

$$j(x, \alpha_R) = \frac{e\hbar}{2mi}[\Psi^\dagger \frac{\partial \Psi}{\partial x} - \frac{\partial \Psi^\dagger}{\partial x}\Psi] + \frac{e\alpha_R}{\hbar}\Psi^\dagger \sigma_z \Psi, \tag{27}$$

for the current density at the semiconductor region of a F/S/F heterojunction including the spin-orbit coupling. We proceed now to discuss the numerical results for the spin-transport properties in a F/S/F heterojunction.

## V. RESULTS AND DISCUSSION

### A. Results at Zero Temperature and without Schottky barriers

We present first the ballistic quantum mechanical results of the spin-transport properties in the absence of Schottky delta-barriers at the interfaces of a F/S/F structure at zero temperature; in the next subsection the case with finite temperature and Schottky barriers will be considered. Fig 2(a) shows the normalized change in conductance [magnetoconductance $\eta$, as defined in Eq. (6)], plotted against $k_R/k_o$ ($k_o \equiv 1 \times 10^5\, cm^{-1}$) for a F/S/F structure with a semiconductor channel length $l_s = 1.0\mu$m (separation between the ferromagnetic contacts) at zero temperature. The effective masses were set to $m^*_f = m_e$ for the ferromagnetic metals, and $m^*_s = 0.036 m_e$ for the InAs-based semiconductor. For the ferromagnets the Fermi wave vectors were set to $k_{F\downarrow} = 1.05 \times 10^8\, cm^{-1}$ and $k_{F\uparrow} = 0.44 \times 10^8\, cm^{-1}$ appropriate for Fe. Note that the same values for the effective masses as well as for the Fermi wave vectors at the ferromagnetic contacts are maintained throughout the paper. The conduction band structure mismatch between the ferromagnet and semiconductor materials was set here to $\delta E_c = 2.0$ eV. An oscillating behavior in $\eta$ is seen as the Rashba spin-orbit coupling strength $k_R$ is varied. If we were to interpret $\eta$ 'semi-classically', *i.e.* as an indicator of spin-injection, the maximum in $\eta$ at $k_R = 1.5k_o$ would signal that the largest amount of electron spin-injection is occurring at this paricular value of $k_R$, Fig.2(a). However Fig. 2(b) shows exactly the opposite, since at resonance ($k_R = 1.5k_o$) an equilibrium condition of the spin-transmissions is reached ($T^P_\uparrow = T^P_\downarrow$ and $T^{AP}_\uparrow = T^{AP}_\downarrow$), hence *no* net spin current is expected to flow through the structure, despite the pronounced spin valve effect seen in Fig. 2(a) at $k_R = 1.5k_o$. This is more clearly shown in Fig. 2(c) where we plot the normalized spin current for the parallel and anti-parallel orientations of the magnetic moments of the ferromagnets. A null result is obtained for $\Delta j^{P/AP}$ at $k_R = 1.5k_o$, which is an equivalent way of saying that no electron spin-injection is taking place. We call this phenomenon the *Quantum Spin Valve* (QSV) effect,[37] since, unlike its familiar classical analog, a non-zero $\eta$ signal can be picked-up in a spin-valve geometry whereas a *zero* net spin current is flowing through the heterostructure. Its origin is inherently due to the coherent quantum interference nature of the spin-transport. Observe that, although $\Delta j^{P/AP} = 0$, in Fig. 2(c) the quantity $2(j^{AP}_\uparrow - j^P_\downarrow) \neq 0$ at $k_R = 1.5k_o$, that is, condition (b.3) of Sec. III is clearly satisfied. Therefore this situation is consistent with Eq.(10) which tells us that a finite value for $\eta$ should be expected (as seen in Fig.2(a)), despite having no net spin-current. We note in passing that condition (*b.1*) is always fulfilled in the $k_R$ range shown in Fig. 2 with the sole exception of $k_R = 1.5k_o$ where condition (*b.3*) is satisfied instead.

The converse effect can also arise, that is, having $\eta = 0$ with a finite electron spin-current flowing in the structure, see Fig.3. Here we have set the conduction band mismatch $\delta E_c = 2.35$ eV, while the rest of the parameters are the same as in Fig.2. Apart from the occurrence of a QSV effect at $k_R \simeq 1.8k_o$, notice that the sign of $\eta$ changes repeatedly as $k_R$ is varied. For instance, at $k_R \simeq 2.9k_o$ the magnetoconductance $\eta$ vanishes, Fig. 3.(a). Therefore, a null spin-injection would be expected, in the standard semi-classical picture. However, at the same value of $k_R$ in Fig. 3(b), there is an imbalance of the spin transmission probabilities since $T_\uparrow^P \neq T_\downarrow^P$ although due to symmetry $T_\uparrow^{AP} = T_\downarrow^{AP}$ always applies. In other words, $\Delta j^P \neq 0$ at that value of Rashba spin-orbit strength $k_R$, as seen in Fig.3(c), which physically means that a net spin-current is in fact flowing when the ferromagnets have parallel magnetization. Notice that the curves for $\Delta j^P$ and $2(j_\uparrow^{AP} - j_\downarrow^P)$ cross each other at this precise value of $k_R$ (Fig.3(c)), which in turns yields the vanishing of $\eta$ for such a spin-orbit strength, fully consistent with Eq. (10) and condition ($a.2$) of Sec. III. Thus we find that in the *coherent quantum regime*, finite spin-injection can occur for the parallel configuration of ferromagnetic electrodes despite $\eta$ being zero, contrary to semi-classical intuition.

In Fig.4 we plot the zero-temperature spin-transmission probabilities as a function of the spin-orbit strength $k_R/k_o$ for two different semiconductor channel lengths, $l_s = 0.1\mu$m, and $1.0\,\mu$m in a F/S/F structure. A wide range of the Rashba spin-orbit strength has been chosen here to better show the strong oscillatory behavior induced by quantum interference as $k_R/k_o$ is tuned. For comparison, the semi-classical ballistic results for the spin-transmission probabilities (Eq.(4) and (7)) have been plotted as well (dotted curves). Notice that the semi-classical curves describe very well the envelopes of the coherent quantum case. Clearly the former do not ever cross, in contrast with the behavior shown in the coherent quantum regime case. From these plots it is clear that a QSV effect appears each time a maximal value of $T_\downarrow^P$ (resonance) is reached as $k_R/k_o$ is swept.

It should be emphasized that the prediction that quantum coherent spin-valve systems may exhibit an unexpected *quantum spin-valve* (QSV) effect does *not* rely at all on the semiconductor-specific Rashba spin-orbit coupling that we include in our model Hamiltonian, but is a general consequence of quantum interference. This is demonstrated in Fig. 5 where the length dependences of the relevant spin-transport parameters are depicted for $k_R/k_o = 0$. The overall behavior of $\eta$, the spin-transmission probabilities, and for the normalized spin-current resemble those studied in Fig. 3 for a fixed channel length. For instance, in Fig. 5(a) at each given maximum in $\eta$, there is a pronounced spin-valve feature, not because of an imbalance (as would be nedeed semi-classically) between $j_\uparrow^P$ and $j_\downarrow^P$, but because the full coherent quantum treatment of spin-transport allows $2(j_\uparrow^{AP} - j_\downarrow^P) \neq 0$ (Fig.5(c)) at the relevant values of $l_s$. This phenomenon, as before, is inherently a *quantum spin valve* effect (but now length dependent) since it is *maximal* where the spin injection *vanishes*, whereas semi-classical reasoning predicts that there should be *no* spin valve effect whenever no spin injection is taking place. We also observe that a finite spin-injection can occur at certain values of $l_s$ whenever a change of sign of the magnetoconductance $\eta$ occurs (Fig.5(a)), *i.e.* even though $\eta$ can be identically equal to zero at those values, which coincide with $\Delta j^P = 2(j_\uparrow^{AP} - j_\downarrow^P) \neq 0$ as stated in condition (*a.2*) of Sec. III.

We have also studied the zero-temperature carrier density dependence of the magnetoconductance $\eta$ in the absence of Rashba spin-orbit coupling ($k_R = 0$), as shown in Fig.6. Here we consider a semiconductor channel of length $0.1\mu$m. To study this dependence we parametrize $k_s$ according to the 2D expression $n_s = k_s^2/2\pi$ for the electron density in the absence of Rashba spin-orbit coupling, where $k_s \equiv k_{F\sigma}^s$ is the degenerate Fermi wave vector in the semiconductor region. Notice that increasing the carrier density at the Fermi energy is equivalent to decreasing the magnitude of conduction band mismatch $\delta E_c$, see Eq. (24). The quantum interference that tunes $\eta$ is exhibited clearly here which produces a strongly oscillatory pattern. Observe that always $|\eta| < 0.1$ for the wide electron density range shown here. For comparison the semi-classical result (Eq. (8)) is also plotted, showing a rather smooth but not monotonic behavior.

### B. Finite Temperature Results with Schottky barriers at the interfaces

The QSV effect described above is predicted to occur at zero temperature and in the absence of potential barriers at the interfaces of a F/S/F heterojunction. We shall now focus on the temperature and Schottky delta-barrier dependence of such QSV phenomena. We begin by discussing the effects of temperature.

We obtain the finite temperature spin-conductances in the Landauer linear response regime through the formula

$$G_\sigma^{P/AP}(T, k_R) = \frac{e^2}{h} \int T_\sigma^{P/AP}(E_\sigma) \left( -\frac{\partial f_D}{\partial E_\sigma} \right) dE_\sigma,$$

with $f_D = \{exp[E_\sigma - E_F)/k_B T] + 1\}^{-1}$ the equilibrium Fermi-Dirac function distribution at the temperature $T$, with the Fermi energy $E_F \equiv E_\sigma(k_{F\sigma}^L)$, such that the splitting energy of the spin-subbands at the ferromagnetic metals is set to $\Delta = \frac{\hbar^2}{2m_f^*}(k_{F\downarrow}^2 - k_{F\uparrow}^2)$. We proceed now to discuss the numerical results for the spin-transport properties in a F/S/F heterojunction at finite temperatures.

In Fig. 7(a) we plot the thermally averaged spin-transmission probabilities in a F/S/F double interface versus $k_R/k_o$ for both magnetization configurations of the ferromagnetic layers, (parallel and anti-parallel) at the temperature of $T = 2.5$ K. The rest of the parameters are the same as in Fig.2. We observe that even at such low temperatures, the effect on the spin-transmission probabilities is quite significant. The feature found at $k_R \sim 1.5k_o$ when $T = 0$ K (see Fig.2) changes qualitatively at such temperatures ($T = 2.5$K) since $T^P_\uparrow \neq T^P_\downarrow$, so that $\Delta j^P \neq 0$ for all $k_R/k_o$, that is, spin-injection that was prevented by quantum interference at zero temperature is now allowed. Therefore the QSV effect evolves towards the standard semi-classical spin-valve behavior as temperature is turned on. Plots of the normalized spin-current for several temperatures show this very sensitive dependence of the signature of the QSV effect on the temperature , Fig.7(b). It is clear that the minimum of $|\Delta j^P|$ smears out very rapidly with temperature; thus the distinctive signature of the quantum spin-valve (QSV) phenomenon is suppressed already at very low temperatures. Similar features are observed if we increase the conduction band mismatch to $\delta E_c = 2.35$ eV, as is shown in Fig.7(c)-(d). We have also plotted the dependence on the relevant spin-transport properties against the semiconductor length (Fig. 8) at a higher temperature ($T = 25$ K), where for clarity we have set $k_R/k_o = 0$. Notice that, although the temperature smearing suppresses the distinctive QSV behavior, since $\Delta j^P \neq 0$ at the maximal values of $\eta$; remarkably however, the converse effect is not degraded at all. We find that in the oscillatory behavior of the magnetoconductance $\eta$ with $l_s$, a vanishing value of $\eta$ is reached always that corresponds to $\Delta j^P = 2(j^{AP}_\uparrow - j^P_\downarrow)$ [see spin-current plots in Fig. 8(c)], despite the relatively high temperature. That is, even though we have a net spin polarized current injected into the structure, it is still possible to have (measure) a zero signal in the magnetoconductance $\eta$. We notice that in fact, the latter effect persists regardless of what the temperature of the system is. We observe also that the amplitudes of all of the oscillations in the plots of Fig. 8 ($T = 25$ K) are very similar to those in Fig. 5 for $T = 0$ K. The physical reason for this is that the energy spacing $\Delta E$ between resonances of the spin-transmission probability plots against Fermi energy (see Fig. 9), are much greater than $K_B T$ at $T = 25$K. The energy spacing $\Delta E$ ranges from 0.01 to 0.04 eV for the relevant Fermi energy interval shown here at zero temperature. Therefore it is expected that even at this relatively high temperature ($T = 25$ K) the thermal smearing will be rather weak, as observed in Fig. 8. It should be noted however that at this high temperature it is very likely that the presence of effects such the inelastic scattering and phase breaking will destroy the coherent quantum interference, and hence the quantum spin-valve effect.

In Fig.10 we show the carrier density dependence of the spin-transport parameters for a F/S/F junction of $l_s = 0.1\mu$m, in the absence of Schottky barriers for two different temperatures, $T = 0$ K, and $T = 25$ K, respectively. The Rashba spin-orbit strength has been set to zero here. The Fabry-Perot like interference pattern of the spin-transmission probabilities caused by the multiple scattering at the interfaces is shown clearly here for a wide carrier density interval. The strong oscillations of the spin-transmission are manifested in a modulation of the magnetoconductance. Even though the curves for $\eta$ look qualitatively very similar when we compare the case with $T = 0$ and case with $T = 25$ K, the smearing effect results in $|\eta(T = 25)| \leq |\eta(T = 0)|$, in general. Notice that the sharp peaks in the transmission probabilities are affected the most by temperature, Fig.10(a) and 10.(c). We point out that the signature of the QSV effect is strongly suppressed here by the temperature (compare Fig. 10.(a)-(b) with Fig. 10.(c)-(d)). However, the quantum interference effect of having a vanishing $\eta$ with a finite spin-current polarization is not affected at all by the temperature, as is seen also in Fig.8.

Finally, we consider the case with delta-potential Schottky barriers at the interfaces of a F/S/F double junction structure. In Fig. 11 the magnetoconductance $\eta$ and the spin-transmission are depicted versus the Rashba spin-orbit strength $k_R/k_o$ for $T = 0$, and $T = 2.5$ K with symmetrical delta-potential (Schottky) barriers of height $U_o$. In the absence of Rashba coupling ($k_R = 0$) we notice that as a result of the introduction of the delta-barriers at the interfaces, $\eta$ suffers a modest enhancement as the height of $U_o$ is increased and $U_o > \Delta$. For finite values of the Rashba spin-orbit strength the finite $U_o$ causes $\eta$ to oscillate in its magnitude [Fig.11(a), and Fig.11(c)]. It is noteworthy that the QSV effect is also found here at $T = 0$ K. Notice that $T^P_\uparrow = T^P_\downarrow$ at $k_R \simeq 3.8k_o$ and $k_R \simeq 4.1k_o$ (Fig.11(b)), whereas $\eta > 0$ at those values of $k_R$. However temperature suppresses the QSV effect since the curves for $T^P_\uparrow$ and $T^P_\downarrow$ do not cross each other in the $k_R/k_o$ interval shown here when $T = 2.5$ K, Fig.11(d). Notice also that the Schottky barriers have a rather strong effect on the spin-transmission probabilities but not on $\eta$, Fig. 11(b) and 11(d). Narrowed peaks in the spin-transmission plots are obtained when the potential barriers at the interfaces are included.

## VI. SUMMARY AND CONCLUSIONS

We have presented a systematic study of the ballistic electron spin transport properties and of the spin-valve phenomena in ferromagnetic metal/semiconductor/ferromagnetic metal structures in the coherent quantum regime of transport. We have investigated in detail the correlation between electron spin-injection and spin-valve behavior. We demonstrated that in the coherent quantum regime the relationship between spin transport and conductance measurements is *qualitatively* different than in the semiclassical regime that has been studied experimentally to date. We have shown in a transparent way that quantum coherence can give rise to a Quantum Spin-Valve (QSV) effect that occurs even in the absence of a net spin current flowing through the heterostructure. We also demonstrated that in the *coherent quantum regime*, the *converse* QSV effect can arise, that is, finite spin-injection is indeed possible for the parallel configuration of ferromagnetic contacts, despite of having a *zero* signal of the magnetoconductance $\eta$, contrary to semi-classical intuition. The effects of Rashba spin-orbit coupling, interface Schottky barriers and temperature on the QSV and its *converse* effect were investigated systematically. We found that the distinctive signature of the QSV effect is extremely sensitive to temperature, as it is suppressed already at very low temperatures. However the *converse* of the QSV effect persists in spite of the thermal smearing of the Fermi function, however it still requires quantum interference and so will be destroyed by inelastic phonon scattering at higher temperatures. The presence of tunnel delta-Schottky barriers at the interfaces was found to enhance the spin-injection efficiencies only slightly contrary to the case of diffusive conductors where tunnel barriers can enhance spin-injection more dramatically.[12,18]

Moreover, the QSV effect remains in the presence of potential barriers at the interfaces at zero temperature, since its origin is due to the multiple scattering at the boundaries. The effect disappears however as the temperature is increased.

In conclusion, we have shown that in the quantum regime of transport a comparison of the conductances of a heterostructure with parallel and antiparallel magnetizations of magnetic contacts can no longer be regarded as an unequivocal indicator as to whether or not spin injection is taking place; it should be supplemented by other probes in studies of coherent spin injection. These surprising conclusions do *not* rely on the semiconductor-specific Rashba spin-orbit coupling that we include in our model Hamiltonian, but are general consequences of quantum interference, although temperature can degrade the effect. These effects should be taken into consideration in interpreting spin injection experiments with spin-valve geometries in the quantum regime of transport.

This work was supported in part by CONACyT-México project No. I39351-E, by NSERC and by the Canadian Institute for Advanced Research.

---

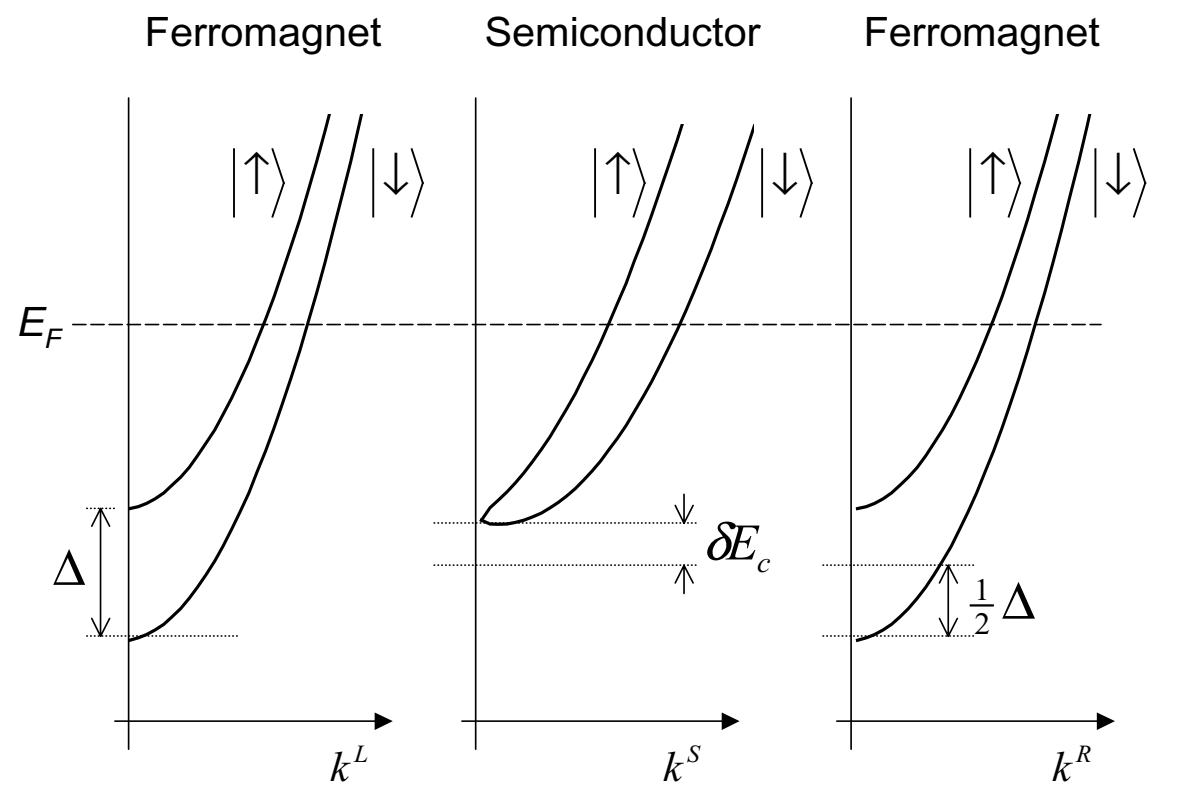

FIG. 1. Schematic diagram of the split bands in a F/S/F heterojunction. The ferromagnet magnetization is chosen to be along the $z$-axis, parallel to the interface. The splitting energy of the spin-subbands of the ferromagnetic metals is defined by $\Delta$, while $\delta E_c$ describes the band structure mismatch between the ferromagnetic and semiconductor materials at the Fermi energy. A finite Rashba spin-orbit coupling is assumed in the semiconductor region which splits the spin-subbands as is schematically shown.

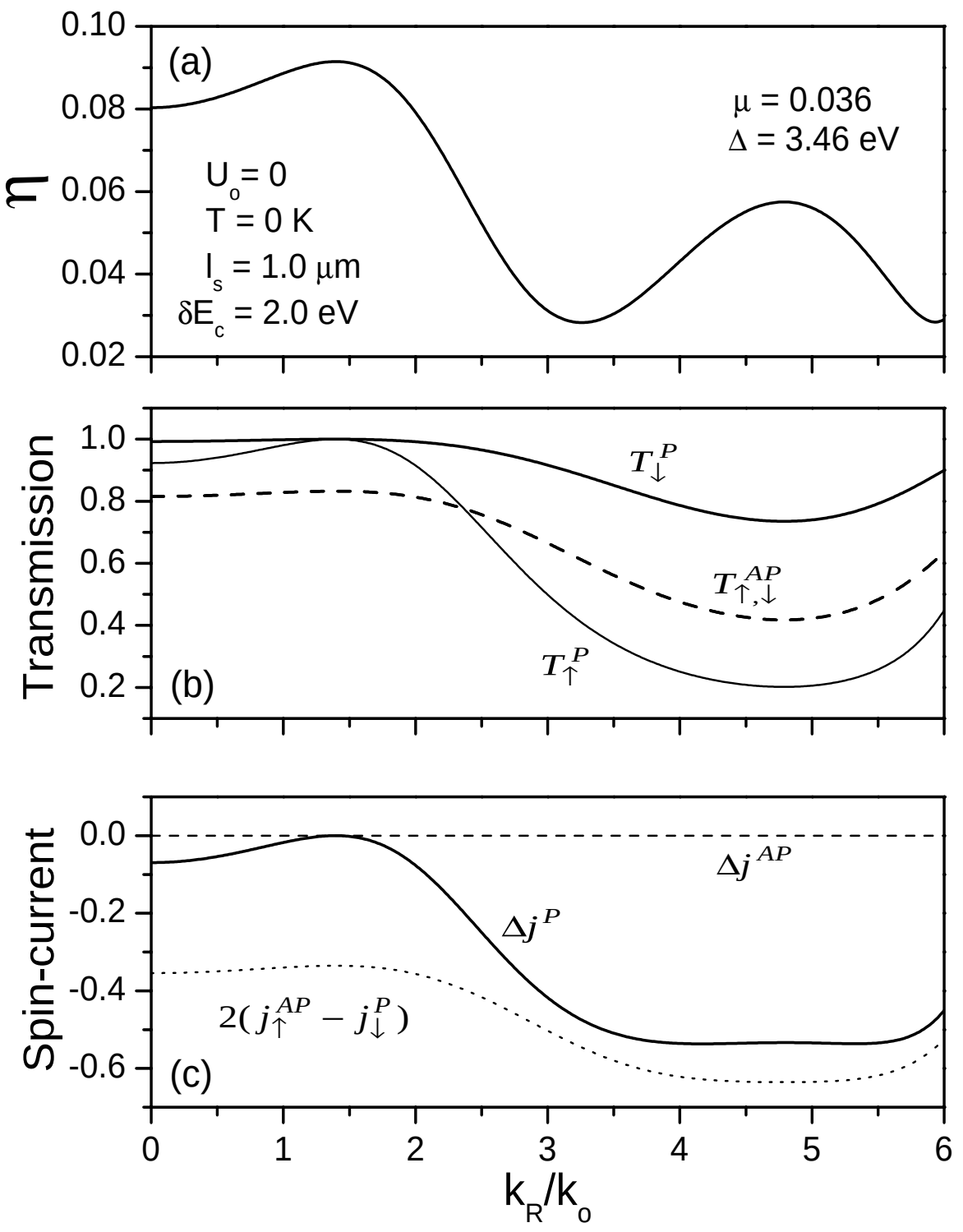

FIG. 2. Zero temperature magnetoconductance $\eta$ (a) spin-transmission probability (b) and normalized spin-currents (c), as a function of the Rashba spin-orbit wave vector $k_R/k_o$ for a F/S/F structure with $l_s = 1.0\mu m$. For the ferromagnets the Fermi wave vectors were chosen $k_{F\downarrow} = 1.05 \times 10^8\, cm^{-1}$ and $k_{F\uparrow} = 0.44 \times 10^8\, cm^{-1}$. The effective masses were set to $m_f^* = m_e$ and $m_s^* = 0.036 m_e$ for InAs. The exchange splitting energy in the ferromagnets has been set to $\Delta = \frac{\hbar^2}{2m_f^*}(k_{F\downarrow}^2 - k_{F\uparrow}^2)$, with a band structure mismatch of $\delta E_c = 2.0\, eV$. Note that at $k_R = 1.5 k_o$ there is maximum in $\eta$ (a) while a zero electron spin-injection is attained at this value of $k_R$, see (b) and (c). This behavior is exactly the opposite of what is expected if $\eta$ is interpreted semi-classically. Due to quantum interference, a Quantum Spin-Valve effect appears, in contradiction with the classical intuition.

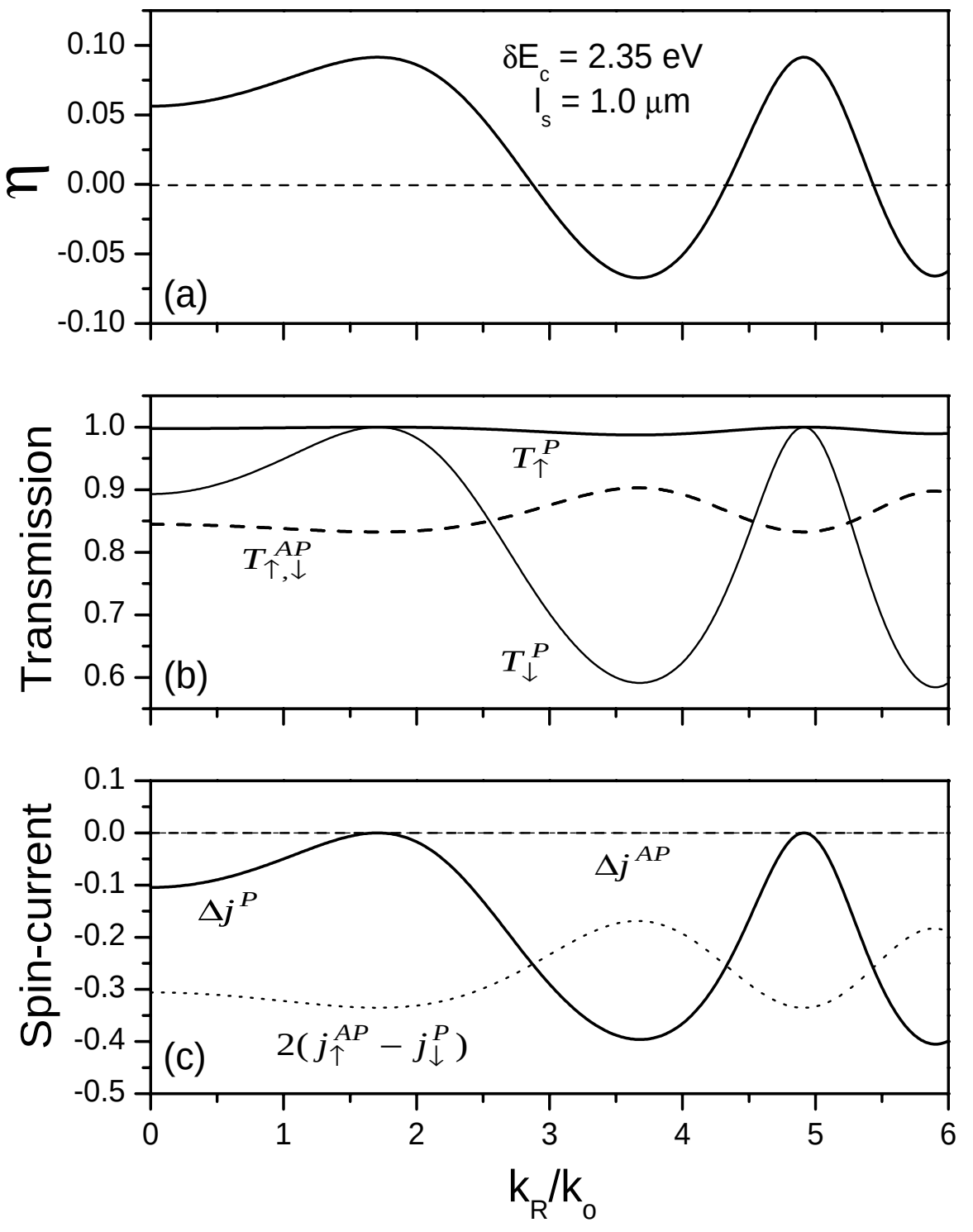

FIG. 3. Magnetoconductance $\eta$ (a) spin-transmission probability (b) and normalized spin-current (c) at $T = 0$, against the Rashba spin-orbit wave vector $k_R/k_o$ for a F/S/F structure with a band structure mismatch of $\delta E_c = 2.0\,eV$, the rest of the parameters are as those in Fig.2. The dashed horizontal line in (a) is to guide the eye. Notice that $\eta$ can change sign as $k_R/k_o$ is increased. A QSV effect is seen at $k_R \simeq 1.8k_o$, similar to that observed in Fig.2. However here the converse effect also occurs. That is, whenever $\eta = 0$, we have $\Delta j^P \neq 0$ (c), which physically means that a finite spin-injection is occurring, contrary again to the semi-classical theory of spin-valve behavior.

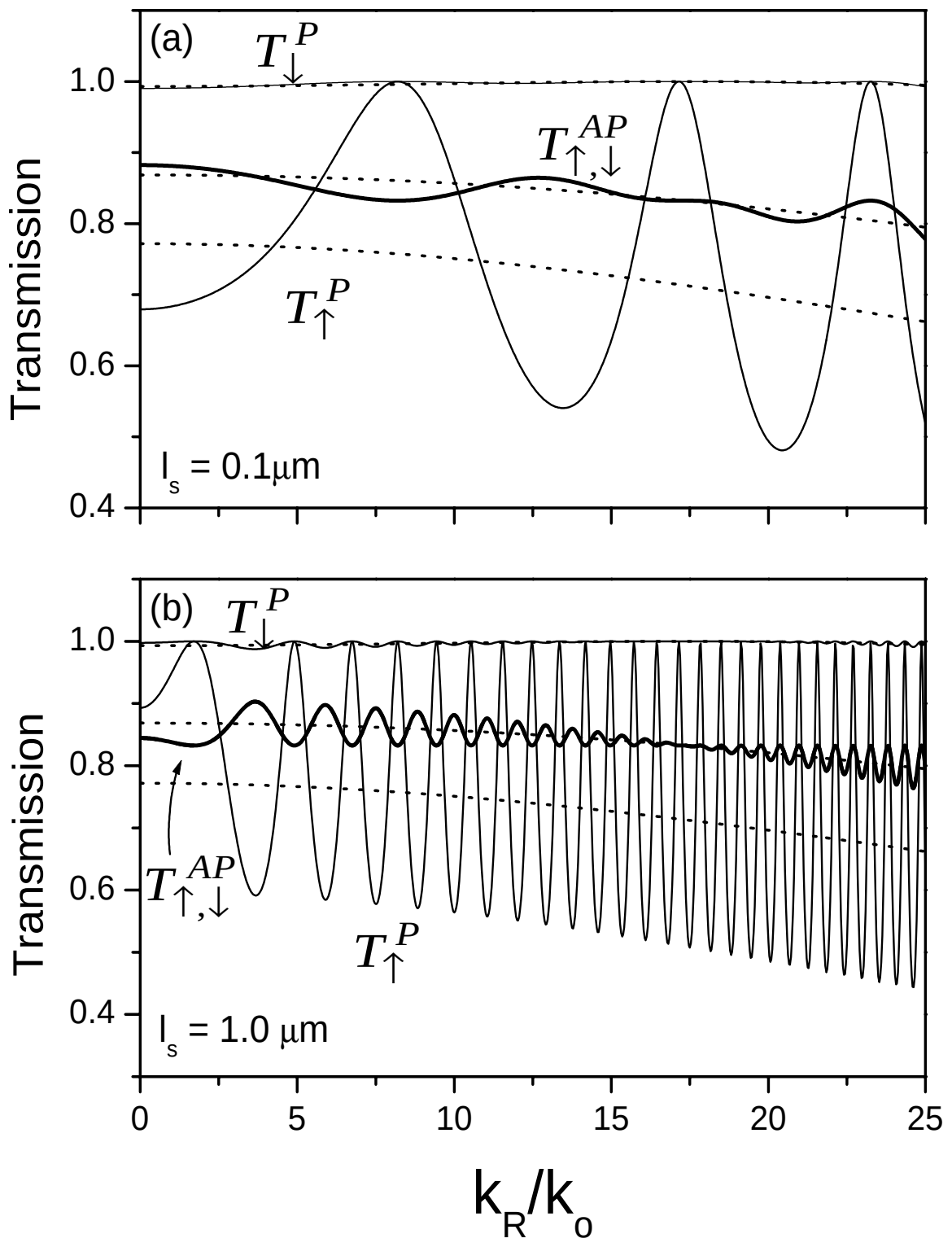

FIG. 4. Spin-transmission probability versus the Rashba spin-orbit wave vector $k_R$ for two different separation length of the ferromagnet electrodes, $l_s = 0.1\mu$m (a) and $l_s = 1.0\mu$m (b), at zero temperature. The rest of the simulation parameters are as in Fig.1. The full quantum treatment for the spin-transport properties gives strong oscillatory features, induced by both the multiple scattering at the boundaries and the tuning in $k_R$. The semi-classical results for each spin-transmission probability have been plotted for comparison (dotted lines).

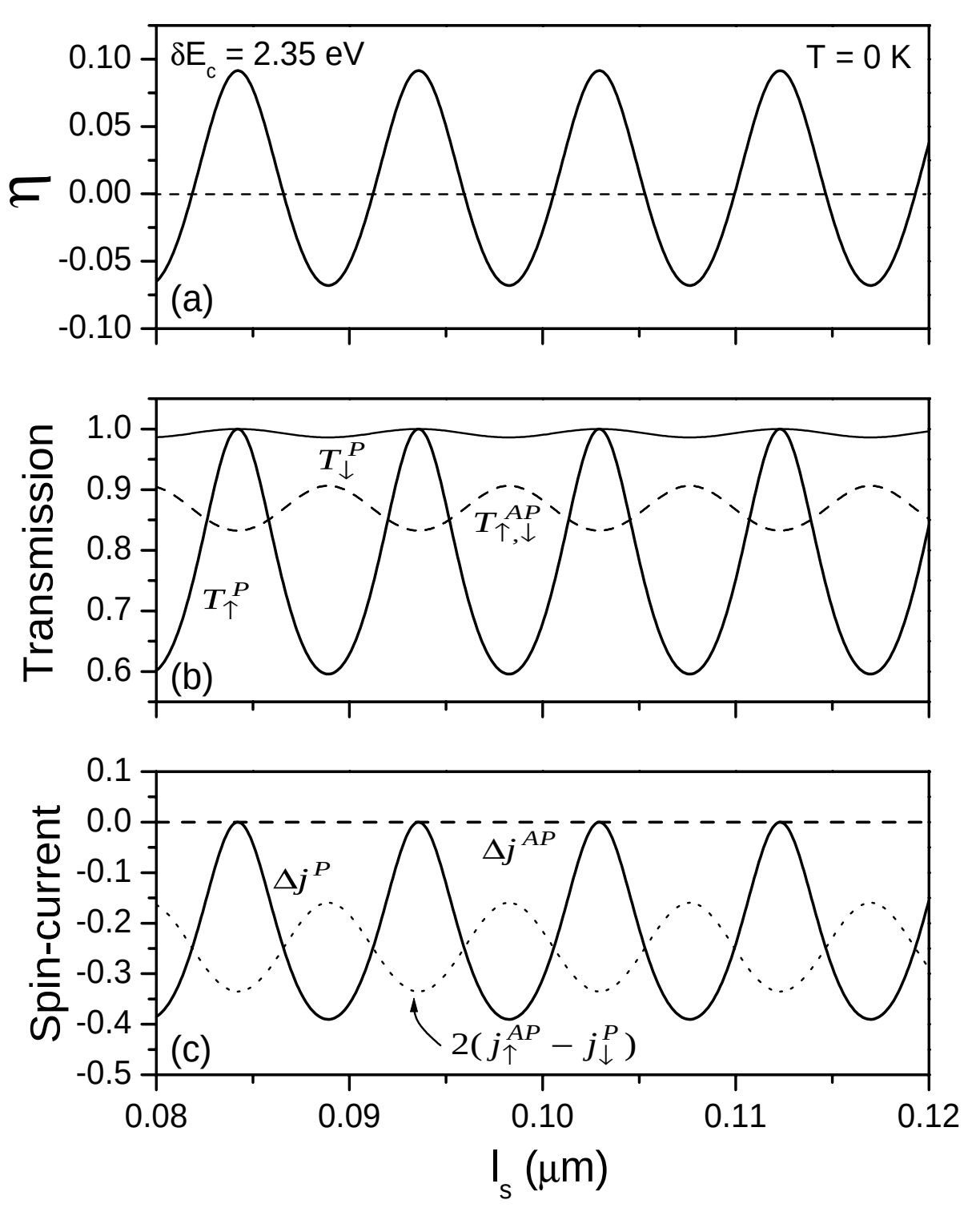

FIG. 5. Length dependence of the magnetoconductance $\eta$ (a), spin-transmission probability (b), and normalized spin-current (c), for a F/S/F double junction with a band structure mismatch of $\delta E_c = 2.35$ eV, and zero Rashba coupling ($k_R = 0$) in the semiconductor region at zero temperature. The dashed horizontal line drawn in (a) at $\eta = 0$ is to guide the eye. These plots show that the origin of the QSV effect does not rely on the Rashba spin-orbit coupling chosen, but it is due to the coherent quantum interference in the F/S/F structure.

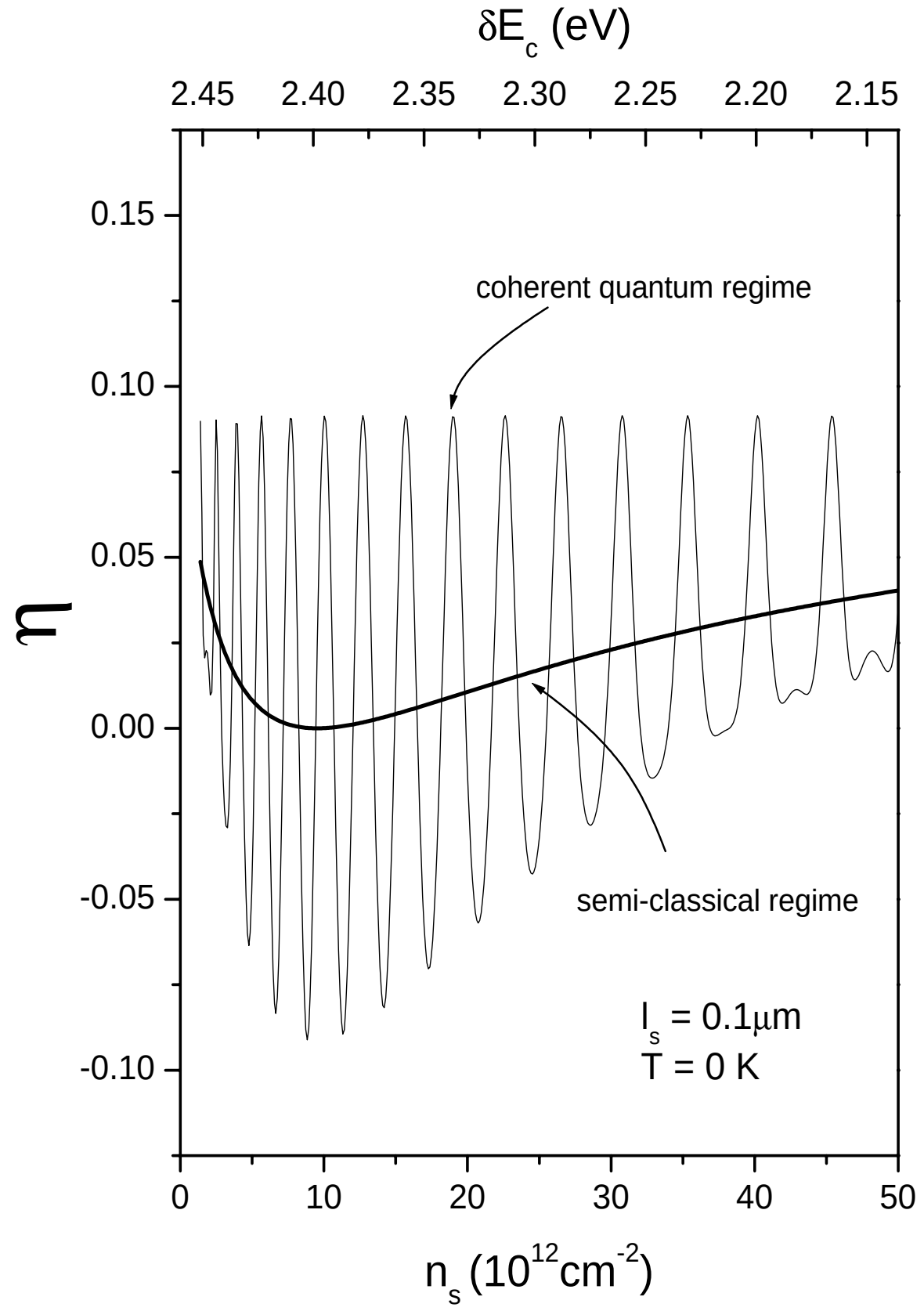

FIG. 6. Magnetoconductance $\eta$ as a function of the carrier density $n_s$ at the semiconductor layer for $l_s = 0.1\mu$m in the absence of Rashba spin-orbit coupling, and at zero temperature. A rather strong oscillatory characteristic is developed as $n_s$ is varied. Clearly in the quantum regime, the magnetoconductance $\eta$ can change of sign at very low carrier densities, in contrast with $\eta$ in the semi-classical regime, which for comparison has been plotted as well (solid thick line). Notice that decreasing the electron density is equivalent of increasing the magnitude of the band structure mismatch energy $\delta E_c$.

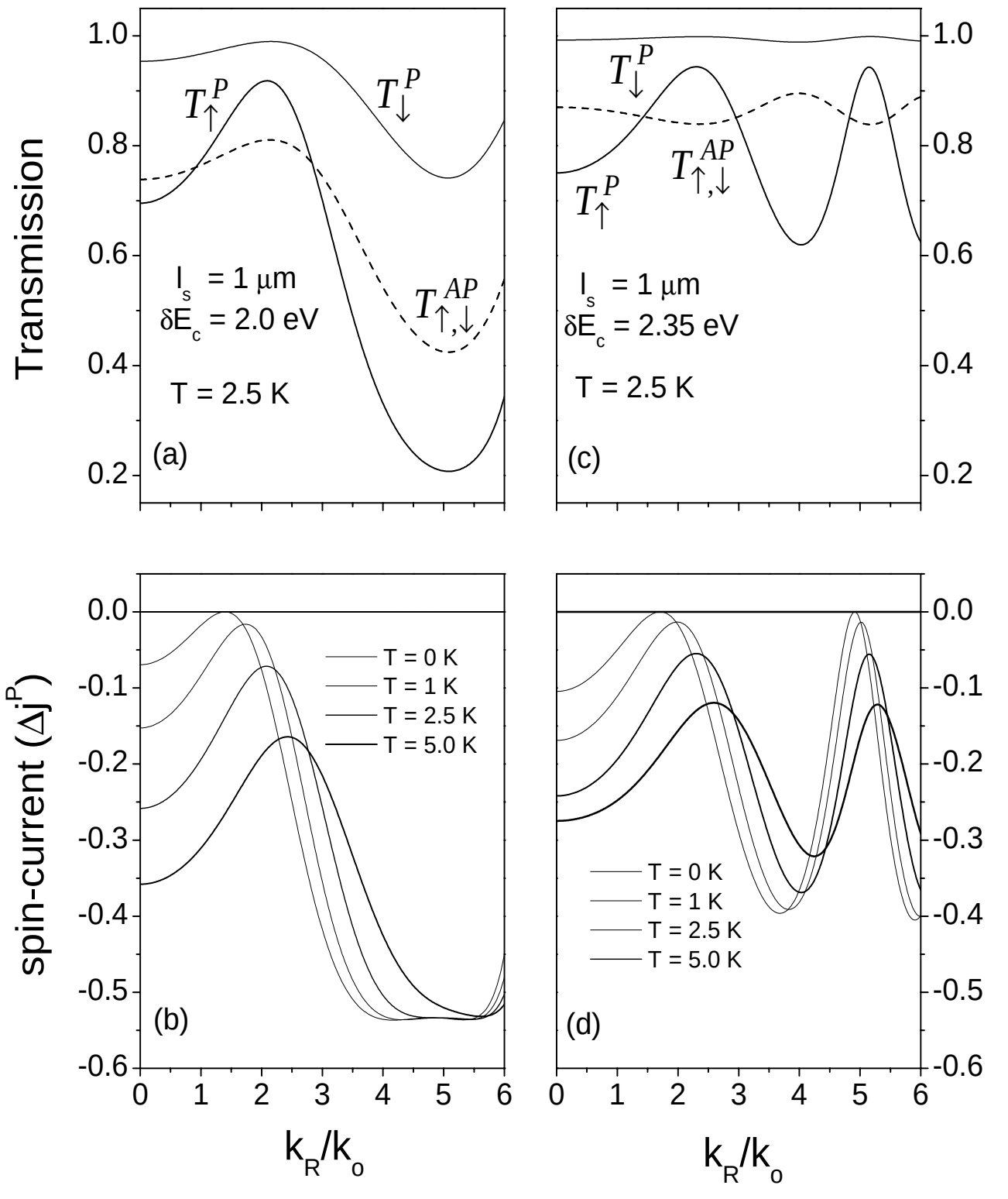

FIG. 7. Spin-transmission and spin-current plots against $k_R/k_o$ at finite temperature for two band structure mismatch energies, 2.0 eV and 2.35 eV, respectively. The distinctive signature of the Quantum Spin-Valve (QSV) effect observed at $T = 0$ (Fig.1) is clearly degraded at the temperatures considered here, $T = 2.5$ K, [(a) and (c)] since $T^P_\uparrow$ and $T^P_\downarrow$ are pushed apart by thermal smearing, yielding $\Delta j^P \neq 0$. In plots (b) and (d) the temperature smearing effect on the spin-current is shown for different temperatures.

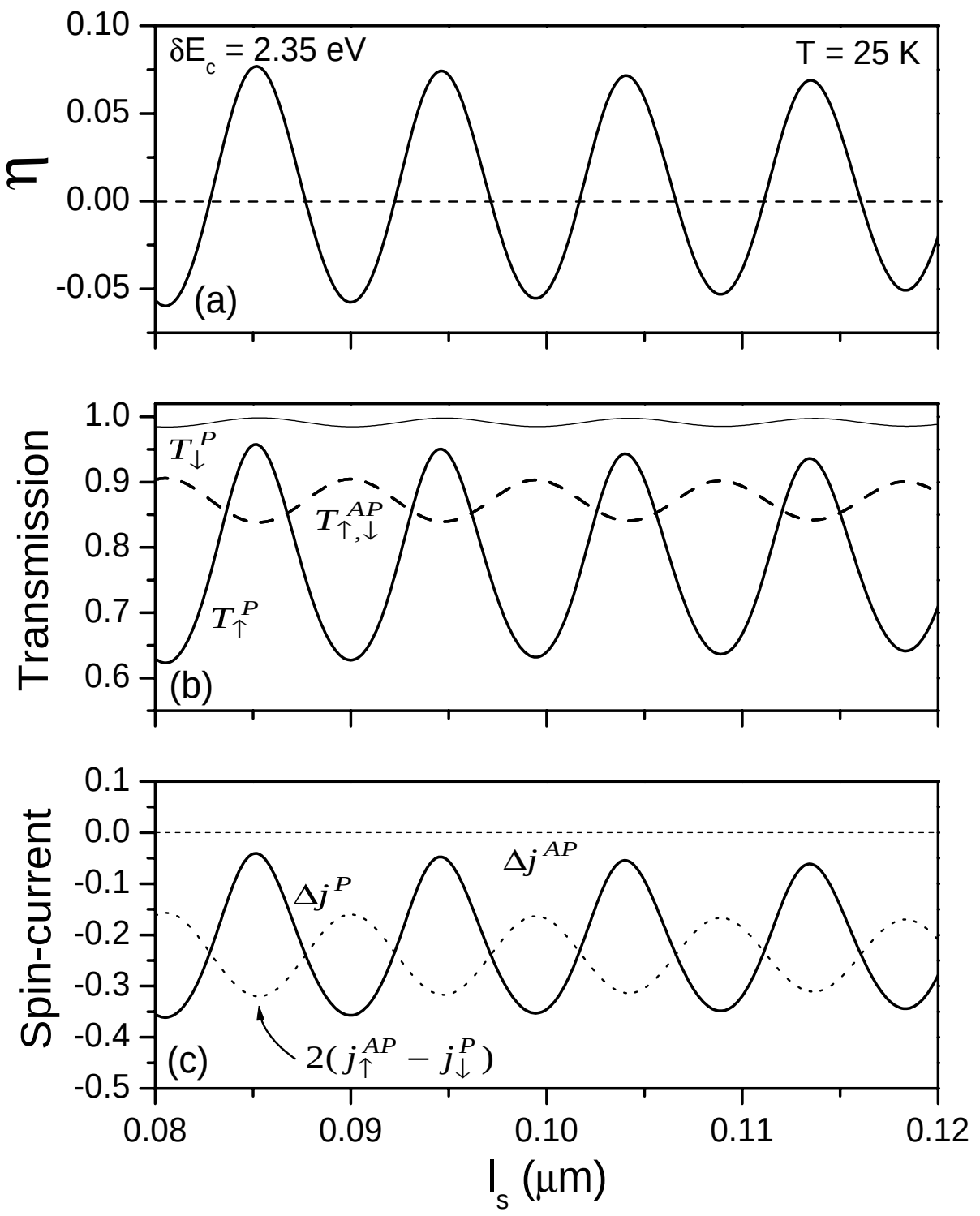

FIG. 8. Finite Temperature length dependence of the magnetoconductance $\eta$ (a), spin-transmission probability (b), and normalized spin-current (c), for a F/S/F double junction with a band structure mismatch of $\delta E_c = 2.35$ eV, and zero Rashba coupling ($k_R = 0$). The dashed horizontal line in (a) at $\eta = 0$ is for guiding the eye. As observed in Fig.7, the QSV effect is suppressed by temperature. However the quantum coherent phenomenon of having $\eta = 0$ for non-vanishing spin-current it is not affected by temperature.

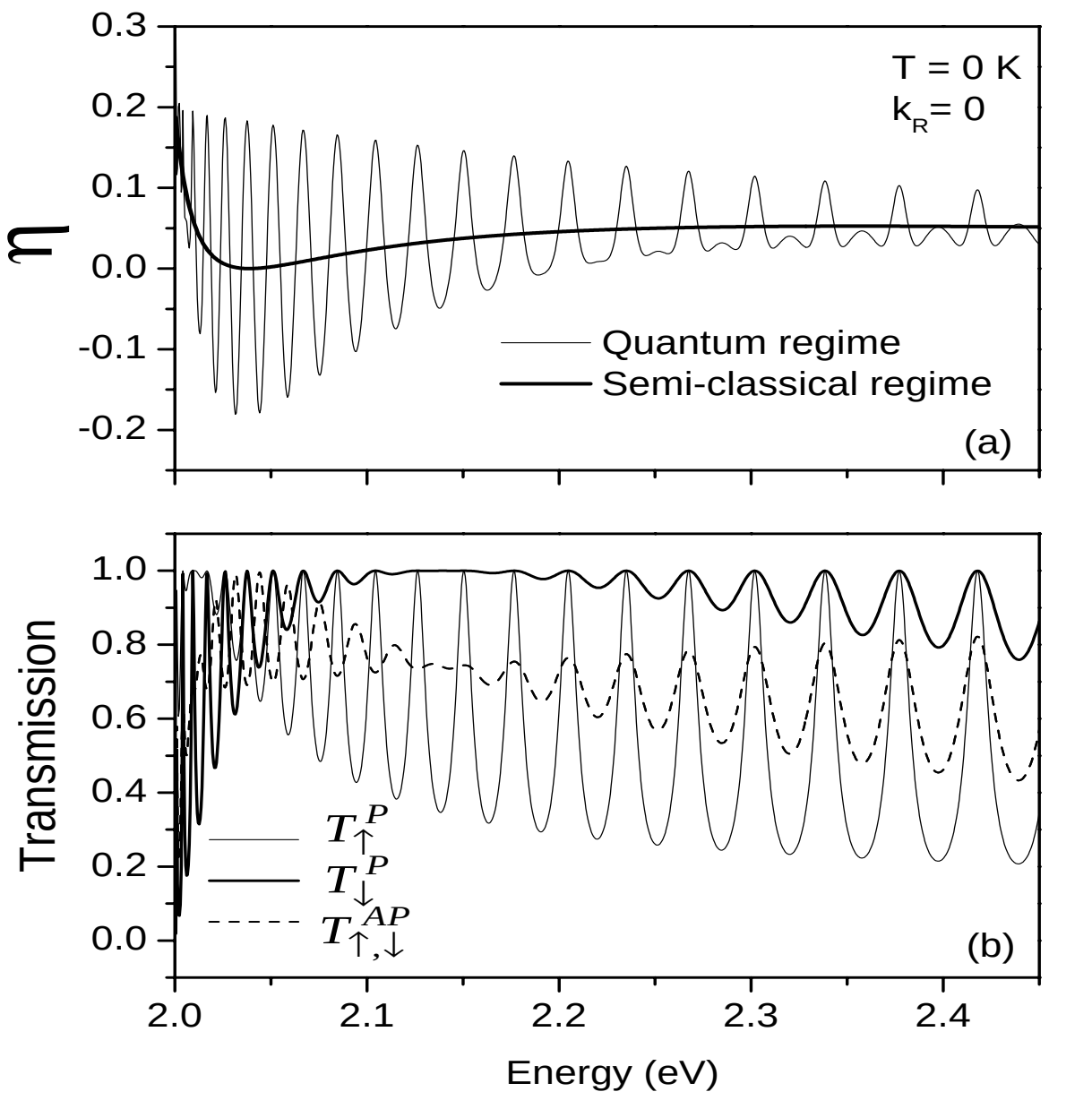

FIG. 9. Semiconductor Fermi Energy behavior of the magnetoconductance (a) and spin-transmission probabilities for a F/S/F junction with a $l_s = 0.1\mu$m without Rashba coupling ($k_R = 0$) at zero temperature. In (a) the semi-classical result has been also plotted for comparisson with the quantum regime. (b) Notice that the energy spacing between resonances $\Delta E \gg K_B T$ at $T = 25$ K.

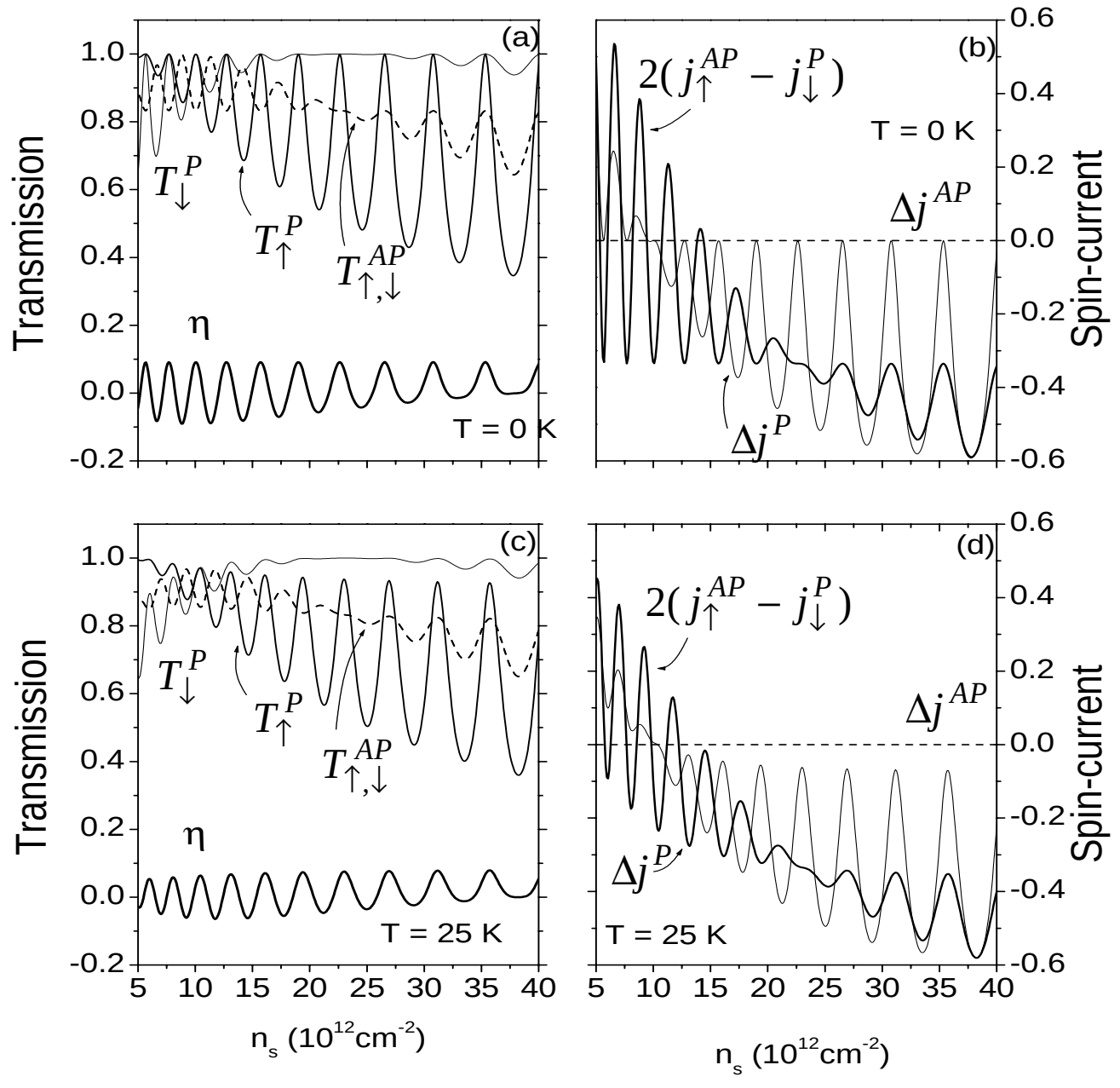

FIG. 10. Carrier density dependence of the magnetoconductance $\eta$, spin-transmission probability, and normalized spin-current at $T = 0$ and $T = 25$ K for a F/S/F junction of $l_s = 0.1\mu$m and $k_R = 0$. The effect of suppression of the QSV signature by the temperature is clearly shown here.

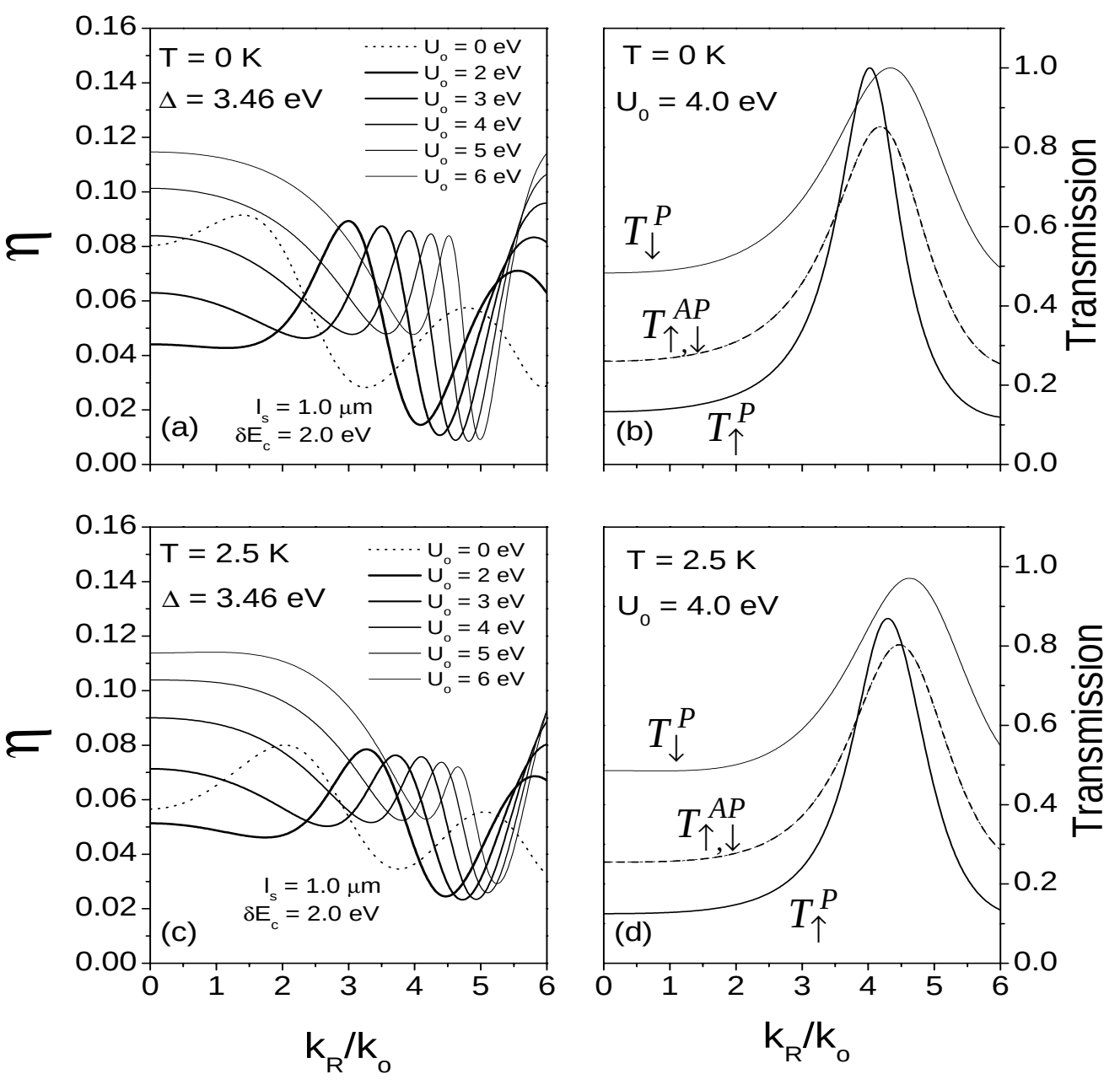

FIG. 11. Magnetoconductance and spin-transmission plots as a function of $k_R/k_o$ at $T = 0$ K and $T = 2.5$ K for F/S/F double junction with delta-Schottky barriers at the interfaces. Here we set $l_s = 1\mu$m and $\delta E_c = 2.0$ eV. At $k_R = 0$ a small but significant enhancement in $\eta$ is observed to occur as the tunnel delta-Schottky barriers ($U_o > \Delta$) are increased in height (a) and (b). At finite $k_R$, $\eta$ is modulated for $k_R > 2k_o$. In contrast, the effect of the Schottky barriers at the interfaces on the spin-transmission probabilities is rather strong (b) and (d). Notice that the QSV effect is also destroyed here by temperature (d).